\documentclass{elsart}
\usepackage{graphicx}
\usepackage{epstopdf}
\usepackage{amssymb}

\begin{document}
% Definitions
\def\nuc#1#2{${}^{#1}$#2}
\def\mee{$\langle m_{\beta\beta} \rangle$}
\def\mnu{$\langle m_{\nu} \rangle$}
\def\gnu{$\langle g_{\nu,\chi}\rangle$}
\def\mmod{$\| \langle m_{\beta\beta} \rangle \|$}
\def\mb{$\langle m_{\beta} \rangle$}
\def\BBz{$\beta\beta(0\nu)$}
\def\BBm{$\beta\beta(0\nu,\chi)$}
\def\BBt{$\beta\beta(2\nu)$}
\def\BB{$\beta\beta$}
\def\Mz{$|M_{0\nu}|$}
\def\Mt{$|M_{2\nu}|$}
\def\Tz{$T^{0\nu}_{1/2}$}
\def\Tt{$T^{2\nu}_{1/2}$}
\def\Tc{$T^{0\nu\,\chi}_{1/2}$}
\def\ms{$\delta m_{\rm sol}^{2}$}
\def\ma{$\delta m_{\rm atm}^{2}$}
\def\ts{$\theta_{\rm sol}$}
\def\ta{$\theta_{\rm atm}$}
\def\tot{$\theta_{13}$}
%end definitions

\begin{frontmatter}



\title{Pulse shape analysis in segmented detectors as a technique for 
background reduction
in Ge double-beta decay experiments}

  \author[label1]{S. R. Elliott,}
  \author[label1]{V. M. Gehman,}
  \author[label2]{K. Kazkaz,}
  \author[label1]{D-M. Mei,} and
  \author[label3]{A. R. Young}
  \address[label1]{Los Alamos National Laboratory, Los Alamos, NM 87545}
  \address[label2]{University of Washington, Seattle, WA 98195}
  \address[label3]{North Carolina State University, Raleigh, NC 27695}

\begin{abstract}
The need to understand and reject backgrounds in Ge-diode detector 
double-beta decay experiments
has given rise to the development of pulse shape analysis in such 
detectors to discern
single-site energy deposits from multiple-site deposits. Here, we 
extend this analysis
to segmented Ge detectors to study the effectiveness of combining 
segmentation with pulse shape
analysis to identify the multiplicity of the energy deposits.
\end{abstract}

\end{frontmatter}

\twocolumn

% main text
\section{Introduction}
Zero-neutrino, double-beta decay \linebreak (\BBz) studies are well 
motivated on physics grounds and several recent reviews
make this case. (See for example Refs. 
\cite{Ell02}\cite{Ell04}\cite{Avi04}\cite{Bar04}.) The process of
\BBz\ may occur in certain even-even nuclei, where $\beta$ decay is 
forbidden, but only if the neutrino is a
massive Majorana particle. During this process a nucleus will change 
atomic number by two units
while emitting 2 electrons ({\it e.g.} $^{76}$Ge $\rightarrow$ 
$^{76}$Se + 2e$^-$). Because the rate of this lepton
number violating process is proportional to the square of the 
effective Majorana neutrino mass
there is strong current interest in this decay. With  the electrons 
being the only light particles
in the final state, the sum of their energies is mono-energetic and 
therefore a distinct signature for the decay.
In any \BBz\ experiment the peak, if it exists, will be superimposed
upon a continuum of background. Present experimental limits indicate
the \BBz\  decay rate would be very low even in the larger mass
 double beta-decay experiments now being
planned, making the identification and elimination of background essential.

For the case of $^{76}$Ge, the total energy that the 2 electrons 
possess is 2.039 MeV. This
energy is above most, but not all, the Q-values of naturally occurring radioactive isotopes.
The range of the \BBz\ electrons in a Ge crystal  is no more than a 
couple mm. Therefore, solid state Ge detectors
make a high-efficiency experimental apparatus with excellent energy resolution
for the study of this decay. The signature for a \BBz\ in
a Ge detector is a localized deposit of ionization of 2.039 MeV; that 
is, a single-site event.
In contrast, $\gamma$ rays of a MeV or more tend to multiply
scatter as they interact in a solid resulting in a multi-site event. 
Furthermore, there are potential cosmogenic isotope decays that can 
populate
the \BBz\ region of interest near 2 MeV including
the $^{68}$Ga $\beta^+$ decay and the $^{60}$Co $\gamma$ cascade. All 
of these background processes tend to produce
multi-site energy deposits. This fundamental difference in the energy 
deposition process creates an opportunity
for separating the signal from background.

Ge detectors have been used for several decades in the study of \BBz\ 
beginning with the initial work of Ref.
\cite{Fiorini} culminating with recent state-of-the-art results from 
the IGEX \cite{IGEX}
and Heidelberg-Moscow \cite{HM} experiments. IIn a coaxial Ge detector,
 the shape of the waveform will primarily depend on the radial distribution of the
ionization produced during an event.
Both IGEX \cite{IGEXpsa} and Heidelberg-Moscow
\cite{HMpsa} used pulse shape analysis (PSA)
for the latter fractions of their data sets in order to improve their 
signal to background ratio.
Both experiments used large coaxial Ge detectors ($\approx$2 kg or 80 mm 
diameter) resulting in a
significant background rejection capability.

In addition to PSA, segmented Ge detectors will also have a 
capability to identify multiple-site energy deposits.
Although this technology has not yet been brought to bear for \BBz\, 
it has been used \cite{GRETA}\cite{AGATA}\cite{VET02} to analyze
the multiple interactions arising from a
$\gamma$ ray impinging on a detector to "track" the $\gamma$-ray 
path. These $\gamma$ tracking arrays require very detailed electric field models for each detector 
in the array\cite{KRO96}.  These models
are then used to simulate a library of pulses for comparison to the measured pulse.
This tracking capability utilizes image-charge formation in adjacent segments to improve event
reconstruction and probably provides the highest resolution localization of events in Ge of any
technique in use at present. 

 Recently the Majorana collaboration \cite{MAJ03} has proposed
to use simple PSA in conjunction with segmentation to further reduce 
background levels in \BBz\ without the need for very detailed detector field models. The detector schemes considered
for this proposal include segmenting coaxial Ge detectors along 
either the axial or azimuthal direction. The electronic
signals from the various segments would provide information regarding 
the axial and azimuthal distributions of ionization and
therefore should improve on the radial-only information provided by 
PSA. Although Ge detectors have been produced with
a large number of segments (the Gretina detector has 36 segments), 
the maximum number of segments isn't necessarily the
optimum for \BBz. The cost is greater for finer segmentation and the 
additional detector contacts with their
associated electronic components and cables may
add to the background.

The purpose of this work is to experimentally measure the 
effectiveness of the combination of PSA and segmentation
in reducing the background for \BBz. Although we also discuss the reduction of 
background via a granularity cut (i.e., requiring only one detector in an array 
observe an energy deposit), we focus on the impact of segmentation. Both segmentation 
and granularity contribute
to an array's self-shielding and hence the identification of single-site energy deposits. 
The results of this study will aid in the 
segmentation design to optimally reduce background.

\section{Experimental Setup}
\begin{figure}%[p]
\begin{center}
\includegraphics[angle=0,width=7.5cm] {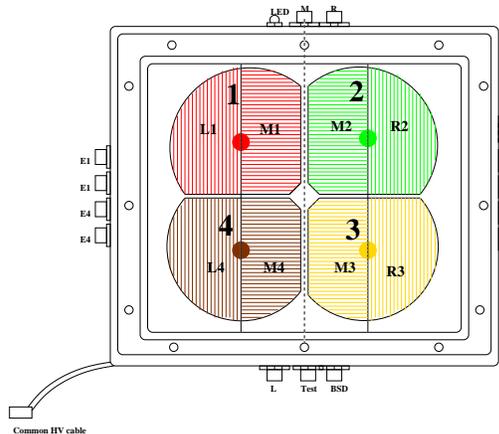}	%{clover_geo_1.eps}
\caption[fig:Clover]{A schematic sketch of the layout of the Ge 
detectors in the CLOVER. This figure is
a rendition from the owner's manual.}
\label{fig:Clover}
\end{center}
\end{figure}

The data for this study were taken using a CLOVER \cite{CloverRef} detector.
The CLOVER is a close-packed array of four germanium detectors, each 
50 mm in diameter and
80 mm long.  Each detector is electrically segmented into
two parts as shown in Fig. \ref{fig:Clover}.  The CLOVER has four 
high-resolution, cold-FET
energy readouts (one for each crystal), and three low-resolution, warm-FET
position readouts corresponding to the left two, middle four and right two
segments.  Coincidences between the energy and position readouts indicate
which segment(s) recorded energy depositions.

\begin{figure}%[p]
\begin{center}
\includegraphics[angle=0,width=7.5cm] {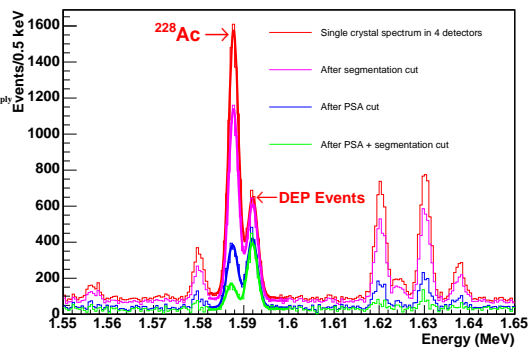}	%{th_data_spect.eps}
\caption{The energy spectrum of events from a Th source near 1.59 MeV.}
\label{fig:spectrum}
\end{center}
\end{figure}

The major focus of PSA is to distinguish
multi-site from single-site energy deposits, and therefore a 
population of each is required for our study.
To this end, we follow the example of Refs. \cite{IGEXpsa} and 
\cite{HMpsa} and use the $\gamma$-ray line from $^{228}$Ac at 1588
keV and the double escape peak at 1592 keV derived from the 2614 keV 
$\gamma$ ray from $^{208}$Tl. (For reference, the mean free path
for a 1.6 (2.6) MeV $\gamma$ ray in solid Ge is about 4 (5) cm.)
Since both of these isotopes are in the decay chain for $^{232}$Th, a 
single source
can be used to obtain both peaks. The 1.6-MeV $\gamma$ ray from $^{228}$Ac typically
interacts about 3 times while depositing its entire energy in a crystal. Hence the full-energy 
$\gamma$-ray peak from $^{228}$Ac consists
almost completely of multi-site events.  The double escape peak 
(DEP), conveniently, is
completely single-site because it requires the annihilation $\gamma$ 
rays from the
electron-positron pair production escape the crystal. A Th source 
spectrum in the energy
region near these peaks is shown in Fig. \ref{fig:spectrum}. Although both of these
processes are required to define the PSA analysis, a 
nearly pure sample of DEP events can be obtained by requiring a coincident 1592-keV
energy deposit in one crystal while simultaneously detecting one of the 511-keV
$\gamma$ rays in a different crystal.

All seven channels from the clover are read out using a pair of X Ray
Instrumentation Associates (XIA) \cite{XIAref}, Digital Gamma Finder 
Four Channel
(DGF4C) CAMAC modules.  These are 14-bit digitizers with a 40 MHz 
sampling rate.
The CAMAC crate is connected to the PCI bus
of a Dell Optiplex computer running Windows 2000.  The system was
controlled using the standard software supplied by XIA.  This data
acquisition software runs in the IGOR Pro environment \cite{IGORref} 
and produces
binary data files that are read in and analyzed using the ROOT framework.

\section{Pulse Shape Analysis}
\begin{figure}%[p]
\begin{center}
\includegraphics[angle=0,width=6.5cm] {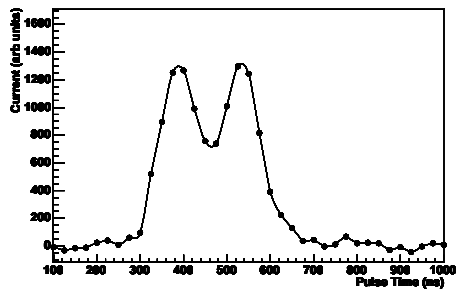}	%{multi_site_pulse-label.pdf}
\caption{The current pulse waveform for a pulse near 1588 keV and 
likely to be an $^{228}$Ac $\gamma$ ray.}
\label{fig:Acpulse}
\end{center}
\end{figure}

\begin{figure}%[p]
\begin{center}
\includegraphics[angle=0,width=6.5cm] {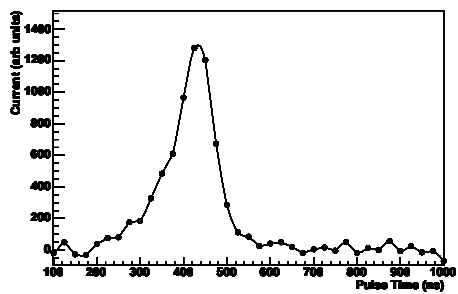} 	%{single_site_pulse-label.pdf}
\caption{The current pulse waveform for a pulse near 1592 keV and 
likely to be a $^{208}$Tl DEP event.}
\label{fig:DEPpulse}
\end{center}
\end{figure}

Figure \ref{fig:Acpulse} shows a current waveform from an event at 
1588 keV and therefore most likely
a multiple-site energy deposit $\gamma$-ray event. Figure 
\ref{fig:DEPpulse} shows a waveform
from an event at 1592 keV and therefore most likely a single-site 
energy deposit from  a
DEP event. The waveforms are clearly different. We analyze the 
waveforms using the formalism
developed by \linebreak Aalseth \cite{AalsethDisertation}.

\begin{figure}%[p]
\begin{center}
\includegraphics[angle=0,width=7.5cm] {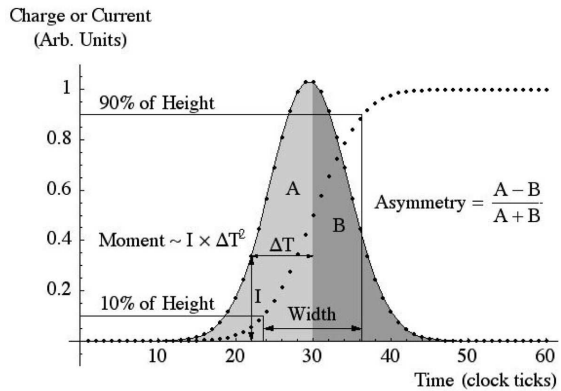}	%{PSDParam.eps}
\caption{Cartoon illustrating idealized pulse shape analysis 
parameters.  The monotonically raising curve
is a charge pulse,
and the shaded-to-zero curve is a current pulse (constructed by 
differentiating the charge
pulse).  See text for a detailed discussion of the calculations of the pulse parametes.}
\label{fig:PulseWidth}
\end{center}
\end{figure}

We calculated three different pulse shape parameters (Fig. 
\ref{fig:PulseWidth})
for each pulse: the width,
front-back asymmetry, and a normalized moment.  The pulse width is defined by
the duration between the times where the charge pulse rises to the baseline plus
10\% and 90\% of the pulse height.  With the pulse mid-point defined as that time 
half way between the 10\% and 90\% points, the front-back asymmetry is the difference
between the area in the front and back halves of the mid-point of the current pulse 
divided by the total
area of the pulse.  Last, the normalized moment 
 parameter is analogous to the
moment of inertia if we treat the current pulse as a mass distribution (i.e.,
it is a sum of the amplitude of the current pulse multiplied by the square of the
distance from the pulse midpoint). The normalization of the normalized moment
parameter ($I_n$) is chosen to minimize any energy dependence. The 
specific formula
  is:

\begin{equation}
I_n = 12\frac{\sum_{i=N_0}^{N}j_i((i - N_{mid})\Delta 
t)^2}{\sum_{i=N_0}^{N}j_i\Delta t}
\end{equation}

where $j_i$ is the current pulse amplitude at time index $i$. $N$, 
$N_{mid}$, and $N_0$ are the time indices at
the pulse onset, midpoint, and end respectively. $\Delta t$ is the 
duration of time
between clock ticks of the digitizer.

\section{Data}

Once we calculate the three PSA parameters, we form a two-dimensional histogram
in the asymmetry and normalized moment parameters.
In our pulse shape analysis, we have tended not to use the pulse width
as an analysis parameter because it is largely degenerate between the
$\gamma$-ray and DEP peak populations. 
Previous studies have found this parameter useful \cite{AalsethDisertation} and we hypothesize
that our smaller crystals lead to less sensitivity to this parameter.

We separately fit empirical functions to the two-dimensional histograms of the
remaining two parameters for DEP and $\gamma$-ray events
in a calibration sample of our data.  The fit results in two
probability distribution functions, one for each class of event 
(single-site or multi-site) that depend upon
the two critical PSA parameters.
For a given event, the measured PSA parameters are used
to calculate the probabilities that it is associated with a given 
class. We assigned the event
to the single-site or multi-site class for which the probability is 
higher. This procedure
results in accepting 75\% of the DEP events but rejects 80\% of the 
$^{228}$Ac $\gamma$-ray events.
The nearby $\gamma$-ray peaks that were not used for this calibration 
have similar rejection results.

One caveat to the comparison between DEP and $^{228}$Ac events is that
they have moderately different spatial distributions throughout the detector. The
DEP events tend to be somewhat nearer the detector edge because both annihilation 
$\gamma$ rays must exit the detector. The mean free path for a 511-keV $\gamma$ ray 
in Ge is about 2 cm, which is smaller than the crystal dimensions. Hence the DEP events tend 
to fall nearer the detector edge where the 511-keV photon has a higher probability to 
escape. In contrast the distribution of the $^{228}$Ac events is defined by the
4-cm mean free path of the $\gamma$ ray to its initial interaction point. In contrast,
to DEP events, \BBz\ events would be uniformly distributed throughout the Ge.

\section{Simulation}
\begin{figure}%[p]
\begin{center}
\includegraphics[angle=0,width=7.5cm] {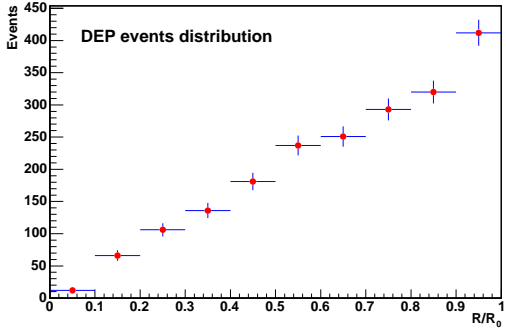}	%{clover_th_pos_dep_r.eps}
\caption{The radial distribution of simulated DEP events for a source 
position in front of the CLOVER. R$_0$
is the radius of the detector (25 mm).}
\label{fig:rsim}
\end{center}
\end{figure}

\begin{figure}%[p]
\begin{center}
\includegraphics[angle=0,width=7.5cm] {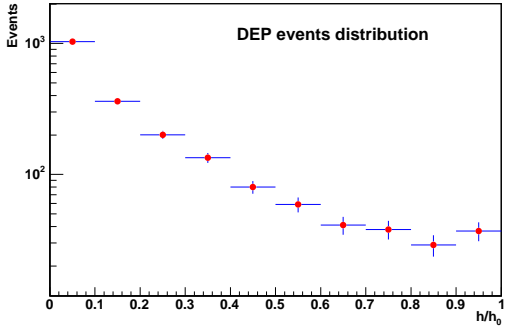}	%{clover_th_pos_dep_z.eps}
\caption{The axial distribution of simulated DEP events for a source 
position in front of the CLOVER. h$_0$
is the total height of the detector (80 mm). The front face of the 
detector is represented by h=0.}
\label{fig:zsim}
\end{center}
\end{figure}

We simulated the distribution of $^{228}$Ac $\gamma$-ray and $^{208}$Tl DEP events 
within a Ge crystal of the
same dimensions as those composing the CLOVER using GEANT3-GCALOR 
\cite{Geant}. We generated r, $\phi$ and
z coordinates for the energy deposits of these events. Figures 
\ref{fig:rsim} and \ref{fig:zsim} show the radial
and axial distributions for the simulated DEP events for a source 
positioned near the front face of the CLOVER.

\begin{figure}%[p]
\begin{center}
\includegraphics[angle=0,width=7.5cm] {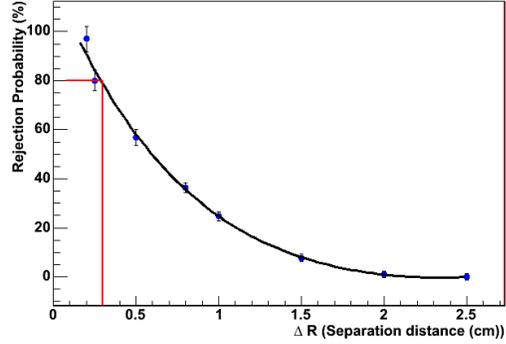}	%{deltaR.eps}
\caption{The fraction of $\gamma$-ray events removed as a function of 
the effective radial separation as predicted
by simulation. Our measured PSA rejection probability indicates an 
effective radial separation of $\approx$ 3 mm for $^{228}$Ac photopeak events
with an average multiplicity of about 3.}
\label{fig:rseparation}
\end{center}
\end{figure}

\begin{figure}%[p]
\begin{center}
\includegraphics[angle=0,width=7.5cm] {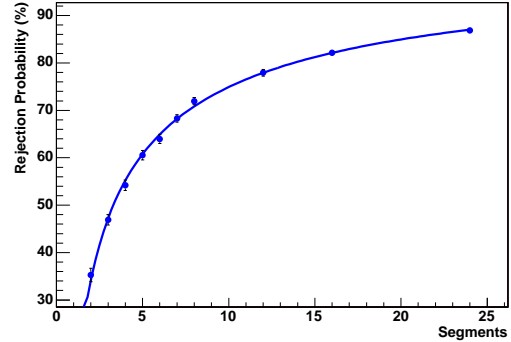}	%{prob_segments_7_23.eps}
\caption{The fraction of $\gamma$-ray events removed as a function of 
the number of segments as predicted by
simulation. The predicted value for 2 segments (35\%) agrees with the measured value
(34\%, see Table 1)}
\label{fig:phiseparation}
\end{center}
\end{figure}

Pulse Shape Analysis effectively eliminates all multi-hit events in which energy is
deposited outside of some "resolution" for identifying radial ionization positions.  This effective
resolution also depends on the multiplicity of the Compton events, improving as the multiplicity
increases (and therefore ensuring the radial ionization distribution exceeds that expected for
single-site events).  To determine the effective resolution, we establish the resolution which
corresponds to rejecting the appropriate number of multi-site events using a Monte-Carlo distribution
for the radial separation between the Compton sites.  This depends on the type of event (photo-peak
vs. Compton continuum) both in the predicted spatial distribution for events and in the resolution.
For the simulated $^{228}$Ac events, we compared our measured survival probability
to the predicted survival as a function of effective separation. From 
these results (Fig. \ref{fig:rseparation})
we estimate
the effective r separation to be 3 mm. That is, an event with 
multiple energy deposits at radii that differ by more
than 3 mm will be interpreted as a multiple-site event. 
Figures such as \ref{fig:rseparation} and \ref{fig:phiseparation} can
therefore be used
to estimate the survival probability for other segmentation 
configurations and PSA spatial resolution
capabilities.

The measurements indicate that about 3\% of the DEP events were rejected
due to the segmentation cut. We determined the number of events before and after
the cut by fitting to the spectrum. Even so, many of these removed events 
might originate from $\gamma$-ray interactions
that form the continuum
beneath the DEP peak. If we assume all the removed events are actually DEP 
events that are excluded because they
fall on the border between the segments, however, we can derive an upper 
estimate for the size of the border. The
border is a diameter across the detector face. Therefore, assuming
the DEP events uniformly illuminated the front face of the detector,
a border width of approximately 1.2 mm would encompass 3\% of the DEP events.

\begin{figure}%[p]
\begin{center}
\includegraphics[angle=0,width=7.5cm] {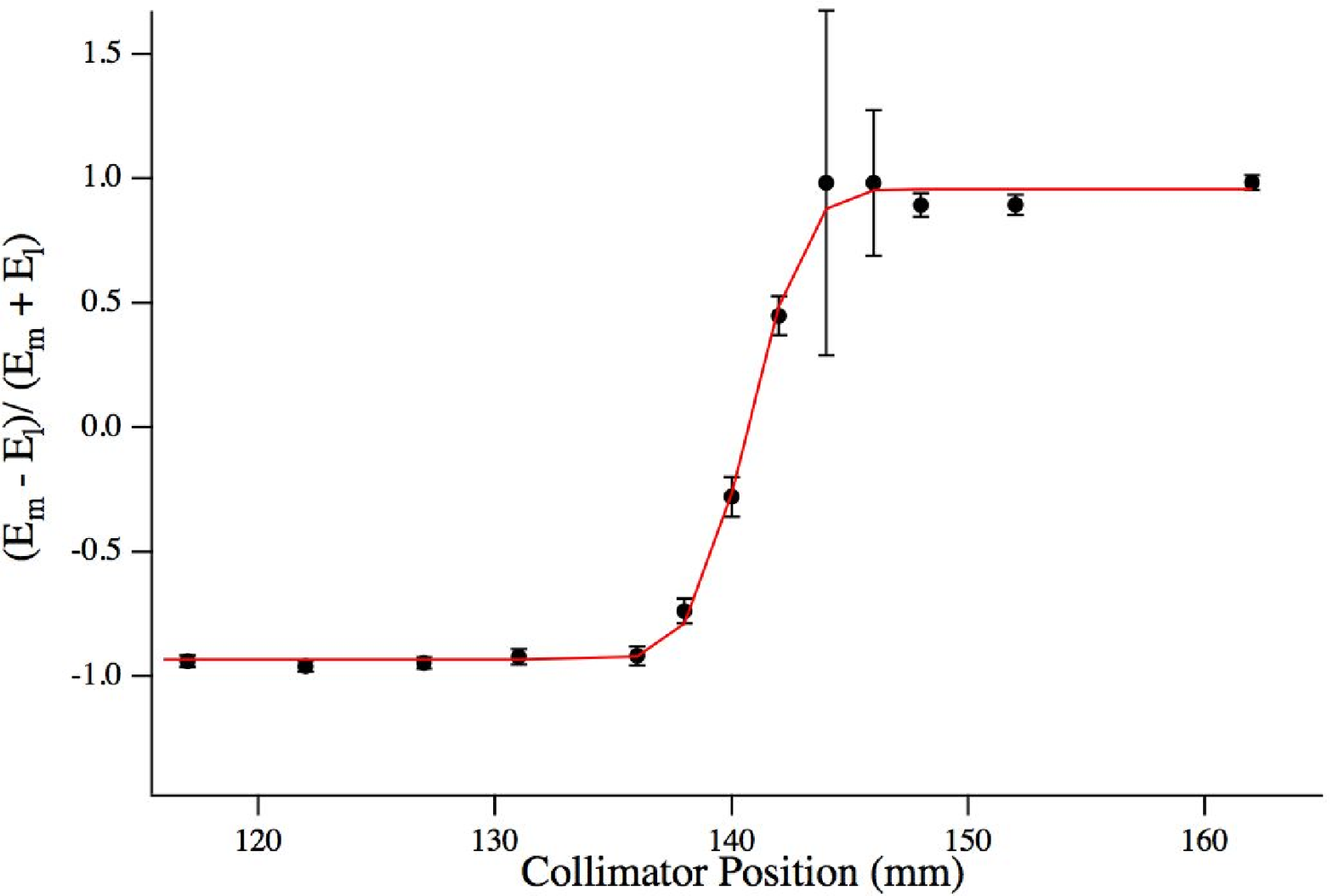}	%{PositionFraction.eps}
\caption{A measurement of the border width of crystal 1 of the CLOVER.
The average fraction of energy assigned to the left and middle 
positions is plotted as a function of source location. The curve is a
fit to the data. See text for
discussion of the collimated source. The absolute value of the 
position scale is arbitrary.}
\label{fig:border}
\end{center}
\end{figure}

To understand the border region better, we scanned the front face of 
the CLOVER with a collimated $^{137}$Cs
\linebreak source. The collimator was formed by a 2-mm hole 
through a 4-inch Pb block. We moved this source across the border in 
steps of 2 mm near the border, and larger steps away from it. We scanned across 
detector number 1 moving from the left
position region through the border region into the middle position 
region. The scanning
was performed slightly off the diameter to avoid the central core of 
the detector.
The collimator illuminated the detector with a 
spot size of 3.5  mm diameter. Fig. \ref{fig:border} shows a plot of the 
average fraction of the energy observed in crystal 1
that is assigned to the left or middle positions for events that 
populate both segments. Specifically we have plotted $(E_m - E_l)/(E_m + E_l)$
where $E_m$ ($E_l$) is the energy in the middle (left) position.
We then fit a step function convolved with a Gaussian to these data and took the 
width of that Gaussian to be the total width corresponding to the finite spot size 
plus the width of the segmentation border.  Since the diameter of the spot 
was known, we could simply subtract it off to obtain the width of the border region.  
We calculated this border region width to be 1.9 mm. That is, 
an event within about 1 mm of the
demarcation line will populate both segments. Since this study was 
done with $\gamma$ rays, most of the two-segment
events are most likely true multi-site events. Therefore this border 
width estimate is also only an upper limit.

\section{Analysis}
\begin{figure}%[p]
\begin{center}
\includegraphics[angle=0,width=7.5cm] {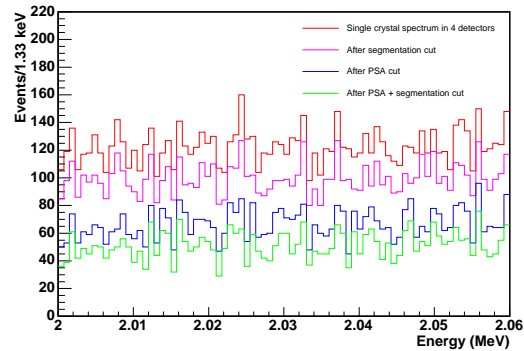}	%{thorium_com_2.eps}
\caption{The energy spectrum of events from a Th source near the 
$^{76}$Ge double-beta decay endpoint of 2039 keV. The top curve is the 
raw spectrum, next is the spectrum after the segmentation cut, next
is the spectrum after the PSA cut, and the bottom curve is after both
segmentation and PSA cuts.}
\label{fig:Compspectrum}
\end{center}
\end{figure}

Figure \ref{fig:Compspectrum} shows the Th source spectrum near the 
$^{76}$Ge double-beta decay endpoint.
Presumably most of the events in this region were from the Compton 
tail of the $^{208}$Tl 2.6-MeV $\gamma$
ray. Such events are a good model of background for double-beta decay 
due to $\gamma$ rays originating outside
the Ge detector.
Events within a full energy peak (such as those from $^{228}$Ac at 
1.58 MeV) are predominantly
multi-site as they are comprised of several $\gamma$-ray 
interactions. The fraction of
events in the Compton tail that are single-site, however, is appreciable. Not 
surprisingly, the rejection of
events in this 2.0 - 2.1 MeV window is less than that for the full 
energy $\gamma$-ray peak. Here the
survival probability for the two cuts is 30\%. In addition, however, a granularity cut
is helpful for such Compton scatters. Requiring that only 1 of the 4 detectors has
a signal above a 30-keV threshold reduces the rate in this window by a factor of $1.89\pm0.023\pm0.09$, where the 
first uncertainty is statistical and the second is a systematic uncertainty estimated by the difference in the results
found from different source placements. All our results concerning the segmentation
and PSA analyses were calculated on data sets after all events failing the granularity test were removed.

Single-escape peak events (SEP) are necessarily multiple-site energy deposits although 
typically with fewer interaction points than a 1.59-MeV $^{228}$Ac $\gamma$-ray event. Thus
the survival probabilities for this class of events is between the $^{228}$Ac $\gamma$ 
ray and the DEP values.

Table \ref{tab:Results} summarizes the survival probabilities for the 
segmentation and PSA cuts and for
the combined cuts. The statistical uncertainties are determined from 
the ratios of events before
and after each cut. The systematic uncertainties are estimated by 
comparing the surviving fractions from
three different data sets, each with the source at a different 
location with respect to the CLOVER. The
source was located near the center of the front face of the CLOVER, 
and near 2 sides of the CLOVER.

\begin{table*}
\caption{\protect  A summary of the survival probabilities for each 
data cut for several processes in the CLOVER. The
first uncertainties are statistical and the second are systematic.}
\label{tab:Results}
\begin{center}
\renewcommand{\arraystretch}{1.2}
\begin{tabular}{lcccc}  \hline\hline
Process                  & Energy      & Segmentation          & PSA 
& Seg. and PSA         \\ \hline
$^{228}$Ac $\gamma$ ray  & 1.588 MeV   & (66$\pm$ 1.2 $\pm$ 0.7)\%  & 
(20$\pm$ 0.5 $\pm$ 1.0)\%  &  (7$\pm$ 0.3 $\pm$ 0.4)\% \\
$^{208}$Tl DEP           & 1.592 MeV   & (97$\pm$ 2.4 $\pm$ 1.2)\%  & 
(75$\pm$ 2.0 $\pm$ 2.1)\%  & (73$\pm$ 2.0 $\pm$ 4.0)\%  \\
$^{208}$Tl SEP           & 2.103 MeV   & (63$\pm$ 1.4 $\pm$ 1.8)\%  & 
(45$\pm$ 1.1 $\pm$ 4.5)\%  & (20$\pm$ 0.7 $\pm$ 2.0)\%  \\
Compton Continuum        & 2.0-2.1 MeV & (81$\pm$ 1.6 $\pm$ 2.0)\%  & 
(43$\pm$ 0.9 $\pm$ 3.0)\%  & (30$\pm$ 0.6 $\pm$ 2.0)\%  \\
$^{208}$Tl $\gamma$ ray  & 2.614 MeV   & (70$\pm$ 2.0 $\pm$ 4.1)\%  & 
(17$\pm$ 0.7 $\pm$ 2.2)\%  &  (8$\pm$ 0.5 $\pm$ 0.4)\%  \\ \hline
\end{tabular}
\end{center}
\end{table*}

The PSA survival results are comparable to those found by Aalseth 
\cite{AalsethDisertation} for the $^{228}$Ac
1.588-MeV line (26\%) and for the DEP line (80\%). The survival 
results are also similar to those of Ref. \cite{KDHK}
(20\% and 62\% respectively).

\begin{figure}%[p]
\begin{center}
\includegraphics[angle=0,width=7.5cm] {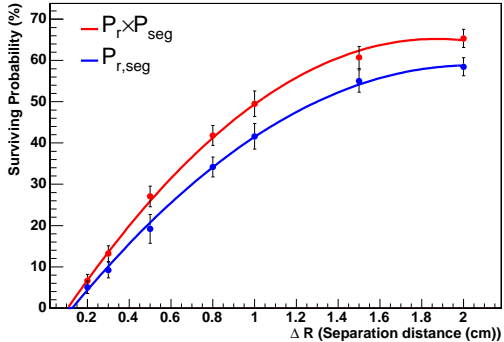}	%{prob_seg_deltaR.eps}
\caption{The joint-cut survival probability (P$_{r,seg}$, lower 
curve) and the product of the individual cut survival
probabilities (P$_{r}\times$P$_{seg}$) as a function of effective 
spatial resolution as predicted by simulation for $^{228}$Ac $\gamma$ rays.
For both curves, the $\phi$ survival is determined from a 
straight-forward segmentation cut. For an effective radial separation of 3 mm, the
simulation agrees with the data.}
\label{fig:ProbDeltaR}
\end{center}
\end{figure}

The data in Fig.
\ref{fig:spectrum} and Table \ref{tab:Results} show that the product 
of the measured survival
probabilities due to pulse shape analysis (P$_{r}$= 0.20)
and that due to the segmentation cut (P$_{seg}$ = 0.66) is greater 
than the measured survival probability
of the joint cut (P$_{r,seg}$ = 0.07). This indicates that the two 
cuts are effective in removing events from
separate, but not necessarily exclusive, subsets of the entire data set.
These measured values agree with the simulation results 
plotted in Fig. \ref{fig:ProbDeltaR}.

\begin{figure}%[p]
\begin{center}
\includegraphics[angle=0,width=7.5cm] {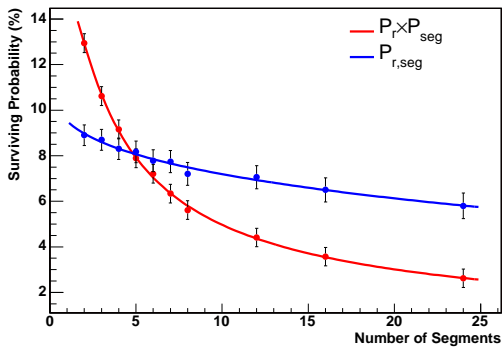}	%{prob_deltaR_seg.eps}
\caption{The joint-cut survival probability (P$_{r,seg}$, the mostly 
higher curve) and the product of the individual cut survival
probabilities (P$_{r}\times$P$_{seg}$) as a function of effective 
azimuthal resolution as predicted by simulation for $^{228}$Ac $\gamma$ rays.
For both curves, the $r$ survival is determined by assigning all 
events with a radial separation greater than 3 mm
to be multiple site.}
\label{fig:ProbDeltaPhi}
\end{center}
\end{figure}

By comparing the simulation to the measurements,
we can develop insight as to how the survival probabilities change
with improvements in the spatial resolutions.  Figs. 
\ref{fig:ProbDeltaR} and \ref{fig:ProbDeltaPhi} show
that, as one would expect, the joint-cut rejection improves as either 
the radial resolution improves or the segmentation increases. For the CLOVER, however,
the spatial resolutions are rather modest. 
The segmentation is coarse and the PSA is
limited due to this modest segmentation. In more highly segmented 
detectors, PSA on the induced pulses from
multiple segments can greatly improve the spatial resolution. Fig. 
\ref{fig:ProbDeltaPhi}
indicates that if the detector's effective resolution is good in
any one dimension, the incremental gain in background rejection by 
improving the resolution in an orthogonal direction
is diminished. In Fig. \ref{fig:ProbDeltaPhi}, the joint-cut 
rejection is shown for a fixed effective
radial resolution (3 mm) but as the angular resolution changes.
The plot shows that as the angular resolution changes
({\it i.e.}, the number of azimuthal segments increases from the Clover's current twofold
azimthal segmentation), a value is reached where
the joint-cut survival probability is no longer better
than the product of the individual cut survival probabilities.

This situation leads to an optimization opportunity. There is no 
fundamental need to increase the segmentation beyond the point
where P$_{r,seg} > $P$_{seg} \times $P$_{r}$. 
Once this condition is 
reached, the added spatial resolution has reached a point of diminishing returns 
with respect to background rejection, but it might be costly to implement. From 
Fig. \ref{fig:ProbDeltaPhi},
this condition is met for an effective radial separation of 3 mm 
and $\approx$5 segments for detectors of the size used here.

\section{Conclusion}
We have measured the rejection of multiple-site energy deposits and 
the acceptance of single-site energy deposits
in a segmented Ge detector by combining pulse-shape analysis with 
segmentation cuts. We have compared
these measurements to simulation to better understand the 
complementarity of the cuts. Since the two cuts operate on
orthogonal coordinates, they tend to remove events from different 
subsets of the data and therefore the
joint action of both tends to be more effective than is predicted 
from the product of each individual cut.
However, as the spatial resolution in one of the dimensions becomes 
better, the impact of the
additional cut is moderated and for very good resolution, the joint 
cut is not as effective as would be predicted
by the product of the individual cuts.

\section{Acknowledgments}
We thank the Majorana collaboration for useful comments and 
suggestions. In particular,
we acknowledge valuable interactions with Craig \linebreak
Aalseth, Reyco Henning, David Radford and Kai Vetter.
This work was supported
in part by Laboratory Directed Research and Development at LANL.

% \appendix

% \section{}
% \label{}

\end{document}